\documentstyle{mn}
\input epsf.sty
\newif\ifAMStwofonts
\AMStwofontstrue
 
\ifoldfss
  \ifCUPmtlplainloaded \else
    \NewTextAlphabet{textbfit} {cmbxti10} {}
    \NewTextAlphabet{textbfss} {cmssbx10} {}
    \NewMathAlphabet{mathbfit} {cmbxti10} {} % for math mode
    \NewMathAlphabet{mathbfss} {cmssbx10} {} %  "   "    "
  \fi
  \ifAMStwofonts
    \ifCUPmtlplainloaded \else
      \NewSymbolFont{upmath} {eurm10}
      \NewSymbolFont{AMSa} {msam10}
      \NewMathSymbol{\upi}     {0}{upmath}{19}
      \NewMathSymbol{\umu}     {0}{upmath}{16}
      \NewMathSymbol{\upartial}{0}{upmath}{40}
      \NewMathSymbol{\leqslant}{3}{AMSa}{36}
      \NewMathSymbol{\geqslant}{3}{AMSa}{3E}

    \fi
  \fi
\fi % End of OFSS
 
\ifnfssone
  \newmathalphabet{\mathit}
  \addtoversion{normal}{\mahit}{cmr}{m}{it}
  \addtoversion{bold}{\mathit}{cmr}{bx}{it}
  \newmathalphabet{\mathbfit} % math mode version of \textbfit{..}
  \addtoversion{normal}{\mathbfit}{cmr}{bx}{it} 
  \addtoversion{bold}{\mathbfit}{cmr}{bx}{it}
  \newmathalphabet{\mathbfss} % math mode version of \textbfss{..}
  \addtoversion{normal}{\mathbfss}{cmss}{bx}{n}
  \addtoversion{bold}{\mathbfss}{cmss}{bx}{n}
  \ifAMStwofonts
    \ifCUPmtlplainloaded \else
and
your
      \UseAMStwoboldmath
      \makeatletter
      \new@mathgroup\upmath@group
      \define@mathgroup\mv@normal\upmath@group{eur}{m}{n}
      \define@mathgroup\mv@bold\upmath@group{eur}{b}{n}
      \edef\UPM{\hexnumber\upmath@group}
      \new@mathgroup\amsa@group
      \define@mathgroup\mv@normal\amsa@group{msa}{m}{n}
      \define@mathgroup\mv@bold\amsa@group{msa}{m}{n}
      \edef\AMSa{\hexnumber\amsa@group}  
      \makeatother
      \mathchardef\upi="0\UPM19
      \mathchardef\umu="0\UPM16
      \mathchardef\upartial="0\UPM40
      \mathchardef\leqslant="3\AMSa36
      \mathchardef\geqslant="3\AMSa3E
    \fi
  \fi
\fi % End of NFSS release 1
   
\ifnfsstwo
  \DeclareMathAlphabet{\mathbfit}{OT1}{cmr}{bx}{it}
  \SetMathAlphabet\mathbfit{bold}{OT1}{cmr}{bx}{it}
  \DeclareMathAlphabet{\mathbfss}{OT1}{cmss}{bx}{n}
  \SetMathAlphabet\mathbfss{bold}{OT1}{cmss}{bx}{n}
  \ifAMStwofonts
    \ifCUPmtlplainloaded \else
      \DeclareSymbolFont{UPM}{U}{eur}{m}{n}
      \SetSymbolFont{UPM}{bold}{U}{eur}{b}{n}
      \DeclareSymbolFont{AMSa}{U}{msa}{m}{n}
      \DeclareMathSymbol{\upi}{0}{UPM}{"19}
      \DeclareMathSymbol{\umu}{0}{UPM}{"16}
      \DeclareMathSymbol{\upartial}{0}{UPM}{"40}
      \DeclareMathSymbol{\leqslant}{3}{AMSa}{"36}
      \DeclareMathSymbol{\geqslant}{3}{AMSa}{"3E}
    \fi
  \fi  
\fi % End of NFSS release 2
      
\ifCUPmtlplainloaded \else
  \ifAMStwofonts \else % If no AMS fonts
    \def\upi{\pi}
    \def\umu{\mu}
    \def\upartial{\partial}
  \fi
\fi

\title{Thermal Conduction and Thermal Instability in the Transition Layer
between an Accretion Disc and a Corona in AGN}
\author[A. R\'o\.za\'nska]
       {A. R\'o\.za\'nska$^1$\\
        $^1$ N. Copernicus Astronomical Center, Bartycka 18, 00-716 Warsaw, 
Poland}
\begin{document}

\maketitle

\begin{abstract}

We study the vertical structure of the transition layer between 
an accretion disc and corona in the context of the existence of 
two-phase medium in thermally unstable regions.
The disc is illuminated by hard X-ray radiation and satisfies the 
condition of hydrostatic equilibrium. 
We take into account the energy exchange between hot corona ($\sim 10^8$ K) 
and cool disc ($\sim 10^4$ K) 
through the radiative processes  and due to thermal conduction. 
We make local stability analysis of the case with conductivity and we 
conclude that thermal conduction does not suppress thermal instability. 
In spite of continuous temperature profile $T(\tau)$ 
there are regions with strong temperature gradient where spontaneous 
perturbations can lead to cloud condensation in the transition layer.
We determine the minimum size $\lambda_{TC}$ of such a perturbation. 
  
\end{abstract}


\section{Introduction}

The X-ray emission from active galactic nuclei (AGN) has been observed
for more than thirty years. 
Although a lot of research were done on this topic in the case of
radio-quiet AGN we still don't know 
neither the mechanism leading to formation of the hot X-ray emitting 
plasma nor the geometry of the X-ray source (for a review, see
Mushotzky, Done \& Pounds 1993, Czerny 1994). 

Recent X-ray data show that a typical X-ray spectrum of a Seyfert 
1 galaxy consists  of a power law 
component with energy index $\sim 0.8-1.0$, high energy 
cutoff above $\sim 200$ keV, and a reflection 
component characterized by low-energy cut-off due to the absorption 
by cold matter, high-energy cut-off due to the Klein-Nishina effect
and the presence of the K$\alpha$ line originating in a cold medium
(e.g. Pounds et al.1990, Zdziarski et al. 1995).

Underlying power-low corresponds to emission from hot region 
and this radiation is most probably thermal 
(annihilation line for par production 
wasn't observed in any spectrum, Macio\l ek-Nied\'zwiecki et al. 1995). 
Second component describes reflection 
of the radiation of hot plasma by cold matter. Approximately half of the
primary emission is observed unprocessed and half of it is reflected.
Fast variability of the line component observed in a number of sources
(Yaqoob et al. 1996) shows that the cold medium is located 
cospatially with the
hot one although some contribution from the reflection by a distant
dusty/molecular torus (Krolik et al. 1994, Iwasawa et al. 1997) 
also might be present.

An attractive scenario of the coexistence of the hot and cold medium
is an optically thin corona above an optically thick accretion disc.
The existence of such a corona was suggested
by Liang \& Price (1977).

Main question about coronae refers to the manner how the energy 
dissipation proceeds in order to  
heat up the plasma to an (electron) temperature $\sim 10^9$K. 
Usually the fraction of energy dissipated in coronae $f$ 
is a free parameter of a model
(Haard \& Maraschi 1991 1993, Kusunose \& Mineshige 1994, Svensson
\& Zdziarski 1994). Another approach is to assume that
the flux in the  
corona is generated by viscosity, similarly like in accretion disc.
Parameterization by viscosity coefficient $\alpha$ allows to predict the ratio
of the energy generated in the disc to that generated in corona. 
Nevertheless in this models full disc/corona vertical structure is 
averaged by two hot and cold layers, and the place of transition is
either adopted arbitrarily  or based on crude approximations
(Nakamura \& Osaki 1993, \.Zycki et al. 1995, Witt et al. 1997).

It is shown (Krolik, Mckee \& Tarter 1981 hereafter KMT) that the 
illumination 
of the cold matter ($\sim10^4$K) by hard X-rays leads to spontaneous
stratification into hot and cold part due to the thermal instability.
The two layers are in pressure equilibrium and the discontinuity in
the temperature and density reflects the discontinuity in the cooling
mechanism: hot layer is cooled by Compton scattering while cold layer
is cooled by atomic processes. This phenomenon underlies the 
spontaneous division of the flow into disc and a corona and therefore
it is important to study it in some detail.

The nature of this phenomenon is well understood due to simple analytical
studies (KMT 1981, Begelman, McKee \& Shields 1983) based on local 
determination of the temperature. Later studies which included the radiative
transfer in the irradiated slab were not suitable for the purpose to 
analyze the problem since either the instability showed up in these codes 
as numerical problems (Raymond 1993) or the iteration had to be started 
below the instability zone (Sincell i Krolik 1997). Therefore we follow the 
basic semi - analytical approach to the description of the effects of 
irradiation but we enrich the physics involved in the process.  

The vertical structure of such a transition between the Compton
cooled and atomic cooled medium in the context  of accretion 
disc/corona system was studied in our previous paper 
(R\' o\. za\' nska \& Czerny 1996 hereafter RC96). 
The model was parameterized by the hard X-ray flux irradiating the transition
layer from above and by the soft UV flux (thermal radiation of the cold disc)
from below. We took into account
the hydrostatic equilibrium and we assumed purely 
radiative energy exchange between the corona and the disc.

The result of that paper suggested that
as soon as there was enough of the hard X-ray emission liberated e.g.
in the upper layer of the corona there was a well defined 
transition
layer between the disc and the corona. In this zone two-phase
medium may form, i.e cool clouds
can coexist with hot plasma under the constant pressure.

In the present paper we discuss the transition layer considering 
also the effect
of thermal conduction. For the first time this effect was studied 
in relation to dissipative corona by 
Macio\l ek-Nied\'zwiecki, Krolik \& Zdziarski (1997). They studied 
the influence of the
heat flux transport on the structure of the corona and 
the transition between the  
hottest layers ($10^9$K) and Compton heated disc atmosphere ($10^8$K).
Their approach did not allow to determine properly the basis of the corona
as they did not include atomic processes and were unable to reach
cool disc layers cooled predominantly by lines and bound-free 
transitions.
Our paper is therefore complementary to theirs as we describe 
more deep parts of the transition starting from 
Compton heated layers and reaching down the cool disc ($10^4$K).

We show that the thermal conduction allows to find very sharp
but nevertheless continuous
solution for density and temperature profile throughout a transition
region, without the need for the two-phase medium. 
However, we study the local stability of such a solution and we show
that this solution is not thermally stable which supports the view
of the existence of the two-phase medium in the transition zone, as
suggested by RC96.    

\section{Assumptions of the vertical structure}

We consider the optically thick and geometrically thin accretion disc
illuminated on the top by hard radiation flux. 
The origin of this hard radiation in radio quiet objects is still unknown 
and usually two models are under debate. 
High energy photons
can rise through the Comptonization: (1) on thermal electrons in 
optically
thin corona above the disc or in the innermost part of the optically
thin disc (2) on non-thermal electrons in the corona or at the basis
of the jet-like (albeit not highly collimated) flow.

We do not specify which interpretation is the correct one. Instead,
we treat the produced hard X-ray flux as external at a given radius.
Therefore,  if the hard X-ray emission is produced in the corona it
means that 
we study the boundary layers between the 
upper, very hot coronal zone and cool disc.      
On the other hand, if the X-ray emission is generated somewhere
in the innermost part of the accretion flow and illuminates the
disc at larger radii  
we actually
calculate the entire corona with the transition zone. To simplify the 
terminology we just call the studied layer 'the transition zone' in both 
cases.  

The incident flux of 
hard radiation $F_{hard0}$ illuminating the upper surface 
of the transition layer is a free parameter in our calculations. It is only 
that part of whole radiation flux generated in X-ray source, which is 
directed
towards the disc. We take into account the 
decrease of this flux by 'absorption on the spot' which  means
full ekstinction of flux through real absorpion and through scattering:   

\begin{equation}
   F_{hard}(\tau)=F_{hard0} \exp (-\tau),
\end{equation}
where $\tau$ is the optical  depth  measured from the top of the slab 
downwards to the cold disc. We don't take into account ekstinction of
flux after one scatter, so the  
matter on considered optical depth does not become the source of
radiation. 
Such an assumption is valid as long as  absorption is the dominant source 
of opacity.
Recent papers on X-ray spectrum reflected from accretion disc suggest 
that in case of illumination 
almost 90\% of incident luminosity is deposited within the cold disc, 
while the remaining 10\% is reflected ( George \& Fabian 1991, Matt, 
Perola \& Piro 1991, Haardt \& Maraschi 1993).
There are strong observational evidences that absorption dominates in the 
illuminated matter (Magdziarz \& Zdziarski 1995).
So most of the X-ray energy is converted into soft radiation and reemited 
according to the formula:
\begin{equation}
    dF_{soft1}(\tau)=F_{hard}(1-a)d \tau
 \end{equation}
 We assume the albedo of hard radiation from the disc atmosphere to
be constant with optical depth and equal $a=0.15$. 

Another source of energy is the gravitational energy of the 
accreting matter itself liberated via the viscous
dissipation (Czerny \& King 1989ab). The soft radiation flux generated 
in this way increases upward throughout the disc 
according to the formula:
 
\begin{equation}
    dF_{soft2}(\tau)=-\frac{3}{2}\frac{ \Omega \alpha}{\rho \kappa_{tot}} 
                    P d \tau,          \label{eq:miekki}                     
 \end{equation}
where $\Omega$ is the angular velocity of keplerian motion,
$ \alpha $ is the viscosity parameter, as introduced by Shakura \& Sunyaev 
(1973), $P$ means the gas pressure, and
$\kappa_{tot}$ is the total opacity coefficient.

Total soft flux emitted by the disc is a sum of two components described 
above:

\begin{equation}
    dF_{soft}(\tau)=dF_{soft1}(\tau)+dF_{soft2}(\tau)     
 \end{equation}

The spectrum of the soft radiation is that of
a black body with the temperature $T_{bb}$ whilst hard X-ray emission is 
assumed to be a power law with the energy index 
$\alpha_E = -0.9$ (here
$F_{\nu} \sim \nu^{\alpha _E}$).
We neglect the change of the spectral shape in any of the two components with
the optical depth.

Strong illumination by X-rays heats up the outer parts of the disc and 
the very hot, optically thin  slab is created above the disc. Such 
a corona generally reduces the radiation pressure, so we assume for the 
transition zone the following equation of state:
 
\begin{equation}
     P=\frac{k}{\mu m_{H}} \rho T,
\end{equation}
with the value of molecular weight $\mu = 0.5$ for cosmic
chemical composition, i.e. we neglect the radiation pressure throughout 
the zone.
 
The vertical distribution of the pressure and the density $\rho$ 
is defined by the
condition of the hydrostatic equilibrium in  vertical direction:

\begin{equation}
     \frac{1}{\rho}\frac{dP}{dz}=-\Omega^2(H_{d}-z)+\frac{F_{soft}}{c}
                     \kappa_{tot}-\frac{F_{hard}}{c}\kappa_{tot} . 
     \label{eq:hydr}
\end{equation}
$H_{d}$ is the half of the disc thickness and 
the coordinate $z$ is measured from the disc half-thickness, $H_{d}$, 
downwards. 

The temperature of a slab is determined by  balancing  radiative heating and 
cooling and heat transport via the thermal conduction from the
neighbor layers. 
Time-independent energy balance equation in vertical direction becomes:

\begin{equation}
  - \rho {\cal L}(\rho,T)=\frac{dq}{dz},              \label{eq:tc}
\end{equation}
where ${\cal L}$ is generalized loss-heat function defined as energy losses 
minus energy gain via the radiative processes
and $q$ is the conductive heat flux.

The solution for  vertical disc/corona structure 
depends on what kind of physical processes are
taken into account. In the hot corona Comptonization and bremsstrahlung
plays main role, but if we want to find the structure of the transition 
layer it is obvious that going closer to the cool, dense disc also atomic 
absorption and emission becomes important. Furthermore, the  accreting 
matter heats itself by viscous forces. 
Also thermal conduction in vertical direction gives substantial 
contribution to the heating/cooling since the transition layer is 
characterized by
strong temperature gradient.
All physical processes which are included in our energy balance equation
are described below in section \ref{sec:pro}.

\section{Energy balance - equilibrium} \label{sec:pro}

\subsection{Radiative processes}
 
Assuming only radiative energy exchange between the layers, the condition 
of thermal equilibrium is satisfied by:
\begin{equation} 
   {\cal L}(\rho,T)=0.
\end{equation} 

Considering following processes: atomic cooling (spontaneous emission, 
bremsstrahlung), atomic heating (absorption in lines and continuum),
Compton heating and cooling and viscous heating, the 
energy equation becomes:
 
\begin{eqnarray}
    \rho\Lambda(T)-F_{hard}\gamma(T)-\kappa_{es} F_{tot}\frac{4k}
 {m_{e}c^{2}}(T_{IC}-T)-  \nonumber \\
     -\frac{3}{2}\Omega\alpha\frac{k}{\mu m_{H}}T=0,  
                \label{eq:bil}
\end{eqnarray} 
where $\Lambda(T)$ and $\gamma(T)$ are cooling and heating functions 
respectively.
We use the same $\Lambda(T)$ and $\gamma(T)$ functions determined by 
CLOUDY code as in our previous paper (RC96) which contains full discussion 
how to obtain these relations. The first function represents the 
total cooling rate 
via the bremsstrahlung, bound-bound and bound-free emission in 
${\rm erg} {\rm cm}^{3}{\rm s}^{-1}{\rm g}^{-2}$ units, the second one 
describes 
the absorption in lines and continuum in the ${\rm cm}^{2}{\rm g}^{-1}$ 
units. 

For radiative processes the energy balance equation is algebraic and we 
can expect
more than one solution for temperature. In fact, we concluded in RC96 that 
the vertical structure of the transition zone
is not continuous, if hydrostatic equilibrium is imposed. For
two narrow ranges in optical depth the matter is described by 
three different values
of temperature and one of them correspond to thermally unstable solution. 
Other two stable solutions cannot be matched continuously. The transition
therefore might be discontinuous and located arbitrarily within 
three-solution
zone (as suggested by Koo \& Kallman 1994) or, as we argue,  
the state of gas is determined by a two-phase equilibrium. 
We concluded 
there was a narrow slab of matter between disc and corona where cool clouds
could coexist with hot plasma under the constant pressure.

To describe the nature of the two-phase medium it is convenient to use the 
ionization parameter $\Xi$ defined after KMT (1981) as:
   \begin{equation}
      \Xi\equiv \frac{F_{ion}}{nk_B Tc},               \label{eq:xibig}
   \end{equation}
where  ${F_{ion}}/{c}$ is  the pressure of the ionizing radiation, $c$ is
the velocity of light, $n$ is the number density of 
particles $[{\rm cm}^{-3}]$ and $k_B$ - Boltzman's constant.

For very large
$\Xi$ only Compton processes are important and the temperature approaches
an asymptotic value (i.e. inverse Compton temperature) 
independent of $\Xi$. With decreasing ionization
parameter bremsstrahlung gradually becomes more important, 
leading to the temperature decrease.
$\Xi^{*}_{h}$ is the minimum value of ionization parameter when high
temperature equilibrium is possible. For lower $\Xi$ bremsstrahlung cooling
overwhelms photoionization heating.

Similar situation is observed for small $\Xi$. The temperature first
increases with increasing ionization parameter (because of increasing of 
ionization level) until ionization
parameter reaches critical value $\Xi^{*}_{c}$. At this point photoionization
overwhelms the line cooling and low temperature equilibrium becomes 
unstable.
 
For narrow range of ionization parameter
\mbox{$\Xi^{*}_{h}<\Xi<\Xi^{*}_{c}$} cold, dense matter can exist in
pressure equilibrium with hot, less dense gas.
Both branches are connected by a third one with the negative slope which 
corresponds to thermally unstable solutions.

The ionization state  of the irradiated gas in thermal equilibrium 
is described as a relation $T(\Xi)$ which depends on:
ionizing flux and the shape of cooling and heating curves $\Lambda(T)$
and $\gamma(T)$. In our situation hard X-ray radiation is predominantly
responsible for the ionization of the gas, and for the eventual presence
of the multi-phase region so we accept
$F_{ion}=F_{hard}$.

\subsection{Thermal conduction}

Trying to solve disc/corona transition we should expect large
temperature gradient so free electrons can efficiently transport the 
heat from upper layers  to the lower ones.
Considering such a situation we solve the equation (\ref{eq:tc})
where $\rho {\cal L}$ contain all radiative processes described above.
 
Usually the conductive heat transport is treated in two limits, 
depending on the effective mean free path of electrons in the medium.
The classical thermal conductivity is based on the assumption that  the 
mean free path is short in comparison with the temperature scale height 
$T/\mid \nabla T \mid$.
For a plasma of cosmic abundance the conductivity is
(Draine \& Giuliani 1984):

\begin{equation}
  \kappa = 5.6\times 10^{-7}\phi_{c} T^{5/2},
\end{equation}
where $\phi_{c}$ is the factor corresponding to reduction 
in the mean free path due to
magnetic fields and turbulences (it is taken to be $\phi_{c}=1$ 
for equal ion and electron temperature).
The heat flux in classical case is expressed by the diffusion
approximation:

\begin{equation}   
  q_{class}=-\kappa \frac{dT}{dz}.                   \label{eq:class}
\end{equation}

When the mean free path is comparable to or greater than the temperature 
scale height the heat flux becomes 'saturated' and nonlocal theory of heat
conduction is required. 
In this situation the heat flux depends on the electron distribution 
function and is limited by two constraints. First is that divergence of the 
current must vanish, and second, that electrons streaming through the ions 
should be stable against various plasma instabilities (like ion acoustic
instability).
To estimate of the flux we rewrite after Cowie and McKee (1977):
 
\begin{equation}
  q_{sat}=5 \phi_{s}\rho c^3=5 \phi_{s}cp,
\end{equation}
where $c^2=p/\rho$ is the isothermal sound speed, $\phi_{s}$ is the 
dimensionless 
uncertainty  parameter. For Maxwellian distribution of electrons
$\phi_{s}=1.1$, but experimental evidences show that usually
$\phi_{s}\sim 0.3$
(Max et al. 1980) and we will use this value in our calculations.

For the purpose of numerical computations, 
it is convenient to have the expression which gives 
smooth transition from  classical diffusive  to the saturated transport.
So we adopt (after Balbus \& McKee 1982) very useful formulae defining 
effective heat flux as:

\begin{equation}
  q=-\frac{\kappa}{1+\sigma}\frac{dT}{dz}.
\end{equation}
where $\sigma$ is the ratio of the classical to the saturated heat flux 
at a given temperature:
\begin{equation}
 \sigma=\left|\frac {q_{class}}{q_{sat}} \right|.    \label{eq:si}
\end{equation}

The energy balance is the differential equation of second order
in this case and we 
expect the temperature to be the monotonic function of the optical depth.
This means the thermal conduction can in principle suppress two-phase 
medium in the
transition zone between disc and corona.  
The temperature profile $T(\tau)$ could be smooth or very sharp.
Smooth one secures stability of the vertical structure solution, but the 
sharp one should be studied more carefully. Perhaps small perturbations of
thermodynamic parameters can lead to thermal instability and
cloud condensation. 

\section{Local Thermal instability}

\subsection{General considerations}

The thermal stability of the matter interacting with radiation field  
was discussed by Field (1965).
He assumed a perturbation of thermodynamic parameters in uniform medium 
which in equilibrium state attains ${\cal L}(\rho_0,T_0)=0$.
He neglected any energy transport and dynamic flows, so his 
isobaric criterion for thermal instability was the following:

\begin{equation} 
 \left(\frac{\partial{\cal L}}{\partial T}\right)_{P}<0,
\end{equation}
where $(\partial{\cal L}/ \partial T)_{P}$ is the isobaric 
evolution of the loss-heat function from the equilibrium. 

Balbus(1986) made more general picture since he took into account the
dynamical perturbations. In this case the loss-heat  function of 
the medium can be a
function of density, temperature and possibly space and time. Thus the 
isobaric instability criterion becomes:

\begin{equation}
 \left( \frac {\partial \ln {\cal L}}{\partial \ln T} \right)_P<1.
\end{equation}

We analyze a local thermal instability taking into 
account the energy transport. Our medium in this case is  
characterized by temperature $T_0 (z)$ and density $\rho_0 (z)$ profile. 
We neglect any dynamic 
flows and we assume that $T_0 (z)$ is  attained by balance
between radiative 
processes and thermal conduction. Our unperturbed medium  
is in stationary state denoted by us as $U(\rho_0,T_0)$ and it satisfies the
condition:

\begin{equation}
   \rho_0 U(\rho_0,T_0)=\rho_0 {\cal L}(\rho_0,T_0)+\frac{dq_0}{dz}=0.
\end{equation}
When total cooling exceeds total heating (via radiative processes and thermal
conduction) then $U>0$.
We perturb two independent thermodynamic  variables (density and 
temperature) under the condition of constant pressure. 
We are interested  only in isobaric 
perturbations because they can lead to clouds condensation.
The specific entropy of the material will change by an amount $\delta S$.
This change evolves in time as:

\begin{equation}
\frac {d}{dt} \delta S=\delta \frac {dS}{dt}=\delta \left(\frac {dQ}
{T dt}\right)=- \delta \left(\frac {U}{T}\right).
\end{equation} 
If $\delta S>0$ then its time derivative should be less than zero to return
perturbed entropy to the background one ( stability). 
For $\frac{d}{dt} \delta S$ with 
positive sign the positive perturbation will grow up from the 
background entropy (instability). Therefore the instability criterion 
follows:

\begin{equation}
 \left[ \frac {\partial (U/T)}{\partial S} \right] _{P}<0.
\end{equation}

Taking into account that in isobaric perturbation $TdS=C_P dT$ and $U=0$ 
for stationary state, our criterion reduces to:

\begin{equation}
\left( \frac {\partial U}{\partial T} \right) _{P}<0.
\end{equation}

\subsection{The perturbation growth rate}

In order to find the possibility of the existence of unstable regions we perform the
local thermal stability analysis. It means
that we assume our medium to be homogeneous locally.
This assumption is done, because our problem
is local. We cannot find full disc structure from the bottom to the disc 
surface for the reason that below our transition layer there is another 
thermally unstable zone as a result of radiation pressure domination 
(Pringle, Rees \& 
Pacholczyk 1973, Shakura \& Sunyaev 1976). 
Global analysis does not allow us to separate this two effects. 
 
At each distance $z$ from the disc surface $z$ we are looking for 
perturbations growing up in time
For this purpose we consider the perturbation of 
two independent parameters: density and temperature, in the form of a wave in 
space and time:

\[  
T=T_0+T_1\exp (ikz+i\omega t), \; \; \; \; \rho=\rho_0+
\rho_1\exp(ikz+i\omega t),
\]
where $k$ is a wavenumber and $\omega$ is frequency. All quantities with 
index $0$ 
characterize unperturbed state. 
 
Field (1965) considered such a perturbation of thermodynamic  parameters
in homogeneous medium and he derived
the critical wavenumber above which the  thermal conduction between 
perturbed element and surroundings suppresses the thermal (radiative)
instability:

\begin{equation}
  k_{c}=\left[ - \frac{\rho_{0}}{\kappa}\left(\frac {\partial{\cal L}}
     {\partial T}\right)_{P}\right]^{1/2}.
\end{equation}

The inverse of critical wavenumber gives  the characteristic length
scale over which perturbation grows up in a result of thermal instability.
Let us to denote this length scale as $\lambda^*=1/k_c$. A $\lambda^*$ 
differs from usually used Field length which is defined as:

\begin{equation}
  \lambda_F=\left[\frac {T\kappa}{\rho {\cal L}_M}\right]^{1/2},  
\end{equation}
where ${\cal L}_M\equiv max(cooling,heating)>0$. 

Physically it means that if perturbed element has a radius $r<\lambda^*$,
then such a cloud will always evaporate due to thermal conduction.
For perturbation with $r>\lambda^*$ cloud can condensate  depending on 
radiative conditions in the surrounding medium (McKee \& Begelman 1990).

For our non-uniform medium with energy transport we still assume only 
isobaric perturbations $\partial P=0$ and we neglect dynamical flows
(so eulerian and lagrangean perturbations are equal).
The matter is characterized by time dependent  energy balance equation:

\begin{equation}
\frac {3}{2} \frac {dP}{dt} -\frac {5}{2}\frac{P}{\rho}\frac{d\rho}{dt}+
\rho {\cal L}(\rho,T) +\frac {dq}{dz}=0.                    \label{eq:per}
\end{equation}
For thermal instability analysis we use the heat flux in classical diffusion
approximation (equation \ref{eq:class}).

The first two terms of equation (\ref{eq:per}) are  $\rho T (dS/dt)$
calculated for perfect gas. 

Considering only linear deviations from the equilibrium we derive 
the relation between $\omega$ and $k$ putting perturbations of density and
temperature  to equations of energy balance and state. 
The growth rate for each value of $z$ is:

\begin{equation}
i \omega= \frac {\kappa_0 k^2+
\rho_0 \left(\frac{\partial U}{\partial T}\right)_P}
{5\rho_0 k_B/m_H}.
\end{equation} 
Here $i \omega$ is always real either negative, or positive. 

We can write temperature deviation as follows:
\begin{equation}
\delta T= T_1 exp(Re(i\omega)t)exp(ikz)
\end{equation}
If the real part of $i \omega$ has positive sign then the perturbation will 
tend to increase, for negative sign perturbation  will be suppressed.
Therefore we can find the critical wavenumber denoted here by $k_{TC}$
which  bounds this two cases. From condition $Re(i \omega)=0$, $k_{TC}$ is 
given by:
 
\begin{equation}
k_{TC}=\left[ -\frac {\rho_0}{\kappa_0}\left(\frac{\partial U}{\partial T}
\right)_P \right]^{1/2}.  \label{eq:ktc}
\end{equation}
This relation reduces to the one of Field (1965) in the case of homogeneous
medium when ${\cal L}_0(\rho_0,T_0)=0$.  
For perfect gas we have:

\begin{equation}
\lefteqn{\left(\frac{\partial U}{\partial T}\right)_P =  \left(\frac 
{\partial{\cal L}} {\partial T} \right)_{\rho}-\frac {{\cal L}_0} {T_0} -
\frac {\rho_0}{T_0} \left(\frac {\partial {\cal L}}{\partial \rho} 
\right)_{T}}  
%\nonumber \\
% & &  -\frac{1}{\rho_0}\left( \frac {\partial}{\partial T}
%\frac{d \kappa}{d T} \right)_\rho \left(\frac {d T_0}{d z}\right)^2- 
%\frac{1}{\rho_0} \left( \frac {\partial \kappa}{\partial T}\right)_{\rho} 
%\frac {d^2 T}{d z^2}.
\end{equation}

The characteristic scale length above which  
the thermal instability develops can be written as $\lambda_{TC}=1/k_{TC}$.
So we expect the  perturbed matter with size $r>\lambda_{TC}$ would condense
in clouds with different density and temperature than the surrounding
medium. 
We guess that  such a perturbation can form a two-phase region in the
non-uniform matter. 

Local linear stability analysis is valid until $exp(Re(i\omega)t)$ is 
small enough
to neglect second order terms. It gives strong constraint
on time over which this analysis applies. The time scale of linear 
perturbation cannot be longer than:

\begin{equation}
t_{max}=\frac{1}{Re(i\omega)}.
\end{equation}

Computation of further evolution would require global stability analysis. This in turn,
would be possible only after the unperturbated full vertical structure is computed. Such 
complete models are not available at present since either surface layers or deep layers are 
ignored in the course of computing (e.g. Sincell \& Krolik 1997, R{\'o}{\.z}a{\'n}ska et al.
1998). The results of local stability analysis cannot therefore describe the final state of 
the transition zone. However they  may show the existence of the thermal instability zone
and strongly indicate the possibility of the formation of the two-phase medium.  

\section{Results}

\subsection{Parameters and Boundary conditions}

To solve the structure of the transition layer we integrate differential
equations presented above as functions of
$\tau$ starting from 
the arbitrary (but low enough) values of the optical depth and density as:
$\tau_{0}=2\times 10^{-5}$, $\rho_{0}=10^{-14}{\rm gcm}^{-3}$.

In a pure radiative case the temperature is determined by solving an
algebraic equation (\ref{eq:bil}) for each optical depth.  
In the case with thermal conduction we integrate second order differential 
equation (\ref{eq:tc}) and the initial
temperature on the top is computed from equation (\ref{eq:bil}).
Second boundary condition is set as the  thermal flux 
being  zero on the surface ($dT/d\tau=0$).

Soft flux generated via viscous dissipation $F_{diss0}$, $F_{hard0}$ and 
X-ray radiation spectrum 
at the disc surface are free parameters in our model.  
We consider in detail a case of a disc emitting soft flux 
$F_{diss0}=5\times 10^{13} {\rm ergcm}^{-2}{\rm s}^{-1}$ which corresponds to
the black body temperature $3.16\times 10^4$ K. Such a flux is expected 
from a 
disc around massive (mass of $10^8 M_{\odot}$) black hole at $10 R_{Schw}$
if the accretion rate within a disc itself is about $2.8\times 10^{24}$ g/s
(i.e. about 0.01 in Eddington units).

We assumed X-ray radiation spectrum with energy cutoff at $200$keV,
as usually observed in Seyfert galaxies (Zdziarski et al. 1995) and with 
spectral 
indexes: $\alpha_E=-0.987$ for $h\nu<200{\rm keV}$ and $\alpha_E=-2.1$
for $h\nu>200$ keV.
Results are presented for different values of hard radiation flux:
$F_{hard0}=5 \times 10^{14}$ and $5 \times 10^{15} 
{\rm ergcm}^{-2}{\rm s}^{-1}$.

In the case without thermal conduction the thickness of the accretion disc
was adopted as a minimum possible value $H_{dmin}$ for which the disc is 
optically thick (RC96). For $H_d < H_{dmin}$ disc 
becomes transparent, whereas for $H_d > H_{dmin}$ disc is no more
geometrically thin and hot atmosphere becomes very narrow in comparison 
with cold matter.
Therefore we adopt the disc thickness $H_{dmin}$ in our approach.
 
Other quantities adopted in our calculations are typical for standard
accretion disc model. We take the  viscosity parameter equal $\alpha=0.1$, 
the Keplerian
angular velocity is 
$\Omega=10^{-5} {\rm s}^{-1}$ (this value corresponds to Keplerian motion
at radius $r=5.1 \times 10^{14}$cm around black hole at the mass of
$10^8 M_{\odot}$), opacity coefficient for electron scattering is
$\kappa_{es}=0.34$ ${\rm cm}^{2}{\rm g}^{-1}$ and Kramer's coefficient
$\kappa_{0}=3.8 \times10^{23}{\rm cm}^{2}{\rm g}^{-1}$.
Total opacity coefficient used in relation \ref{eq:miekki}
is computed from the standard relation \mbox{$\kappa_{tot}=
\kappa_{es}+\kappa_{0}\rho T^{-3.5}$}.

The results are presented as the functions of optical depth $\tau$ or of
the distance from the disc surface $z$. Both parameters are connected
by relation:

\begin{equation}
     dz=\frac{d\tau}{\kappa_{tot}(T) \rho}.       \label{eq:zz}
\end{equation}

\subsection{The vertical structure of the transition layer}

The computations show that in our model heat flux is carried according to
classical description, because $\sigma \ll 1$ (see eq. \ref{eq:si}).

Cool, optically thick disc is almost isothermal so, the thermal conduction 
between the layers is negligible. Similar situation is observed in hot 
corona, but some heat flux can appear there until corona has really 
hight temperature (classical conductivity is proportional to the $T^{5/3}$).
Our numerical results show that heat flux generated in corona is about 
six orders of magnitude smaller than the one in transition layer.

Fig.1. presents the temperature profiles $T(\tau)$ (I) and
stability curves $T(\Xi)$ (II) of the disc illuminated by: a) 
$F_{hard0}=5\times 10^{14}$ ${\rm ergcm}^{-2}{\rm s}^{-1}$
and b) $F_{hard0}=5\times 10^{15}$ ${\rm ergcm}^{-2}{\rm s}^{-1}$. 
Computations for the case when thermal conduction is taken 
into account (solid line) are compared to the ones without thermal 
conduction (points). 
Near the surface, the disc atmosphere  for both cases is heated up to 
the same temperature for given $F_{hard0}$. 
The value of this temperature is close to the Inverse Compton temperature 
which 
is determined by the spectral shape of incident X-ray radiation. 
In our case $T_{IC}=8.19 \times 10^7$K on the top of the discussed  layer.
The zone with hot plasma is thicker for larger $F_{hard0}$.

When the density increases with the optical depth the importance of the  
bremsstrahlung cooling increases and  the temperature starts to 
fall down. For the case with conductivity and for lower $F_{hard0}$ this 
point appears at lower optical depth than for pure radiative case.  

It is shown in previous paper (Figure 5 of RC96) that our method is 
applicable for 
$\tau>1$ and we do not have to consider the radiative transport equation
more carefully.

Going further 
inside the disc temperature reaches the intermediate stable region 
at $T\sim 8\times 10^5$K. 
Since  photoionization is the predominant process in
the thin transition layer between hot $T\sim 10^8$K and cold $T\sim 10^4$K
matter, we can expect that 
heavy elements abundance influences the shape of instability curve. 
For some configuration of elemental abundance even two intermediate stable
regions can exist (Hess et al.1997).          
In our model one intermediate equilibrium region forms and it is wider for 
larger values of X-ray radiation flux.

\begin{figure}
 \epsfxsize = 120mm \epsfbox[50 400 530 700]{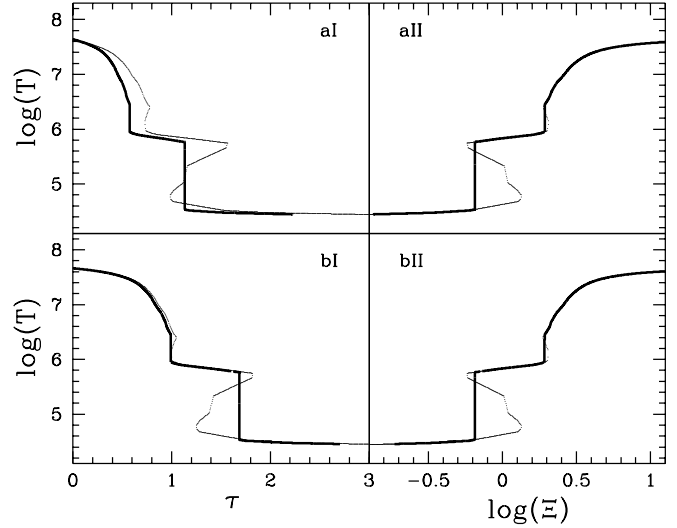}
 \caption{Temperature versus optical depth $\tau$ (left hand side I) and 
          ionization parameter $\Xi$ (right hand side II) for two 
          different values of incident  hard radiation flux: a) 
          $F_{hard0}=5 \times 10^{14}$ ${\rm ergcm}^{-2}{\rm s}^{-1}$, b) 
          $F_{hard0}=5 \times 10^{15}$ ${\rm ergcm}^{-2}{\rm s}^{-1}$.
          Solid wide line represents the case with conductivity, thin dotted
          line shows the pure radiative case.}
\end{figure} 

The most interesting feature of the presented profiles is that 
for the pure radiative case  there are 
two narrow ranges of $\tau$ when matter can reach 
three different values of temperature and one of them  is unstable 
$dT/d \Xi<0$. 
For the case with conductivity the profiles are very sharp but 
continuous opening a possibility that
the thermal conduction suppresses thermal instability in the transition layer 
between disc and coronae. 

Finally, the temperature drops to the value of $2.8\times 10^4$ 
characteristic 
for the cool disc. Bigger flux of X-ray radiation illuminating the surface 
leads to higher optical depth reached by cold matter. 
Our computations for the case with conductivity are stopped at the point 
where the thermal conduction 
flux vanishes at low temperature equilibrium. 

Figures named by II (right hand side) show ionization curves and we can 
easily see that 
thermal conduction does not change the ionization state of matter in regions
of thermal stability. 
But two discontinuous parts of these lines which are characteristic for the 
pure radiative case  are replaced by ones with positive slope $dT/d\Xi>0$. 
So the solution of the disc structure for stationary state is very steep,
but monotonic.  

Fig.2. presents variations of other disc parameters like: pressure (I),
density (II), and  distance from the surface $z$ (III) with optical depth. 
We again compare pure radiative case  (points) and the case with 
conductivity (line)  for two 
values of $F_{hard0}$ a) and b) which are described above.

Near the surface the profiles are identical for both  cases.
At some optical depth (different for each parameter)  lines start differ 
and when thermal conduction is 
taken into account gas attains higher pressure and density at lower 
distance from the surface than without the thermal conduction. 
This difference is bigger for lower value of the initial hard radiation 
flux a).

\begin{figure}
 \epsfxsize = 120mm \epsfbox[50 300 530 700]{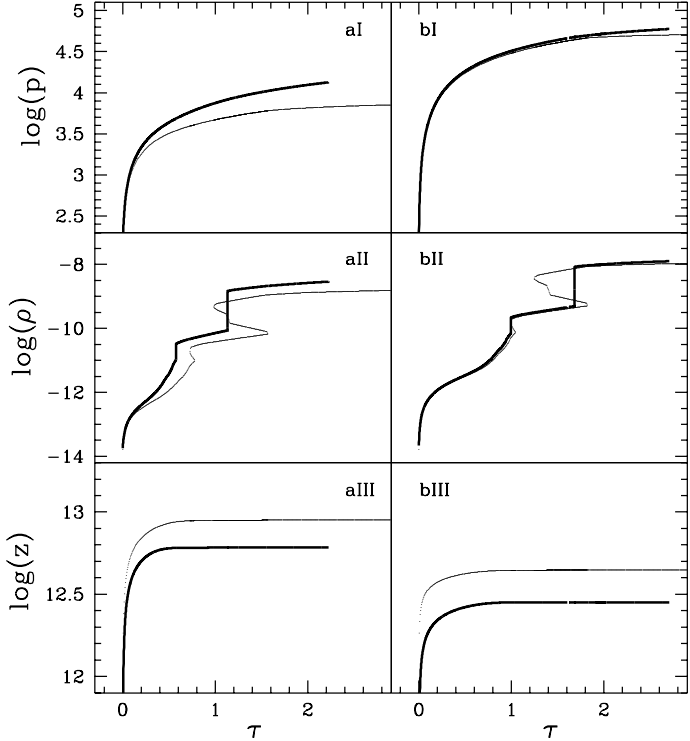}
 \caption{Vertical distribution of I) pressure, II) density and III) optical 
          depth at two different values of incident  hard radiation flux: 
          left hand side a) 
          $F_{hard0}=5 \times 10^{14}$ ${\rm ergcm}^{-2}{\rm s}^{-1}$, 
          right hand side b) 
          $F_{hard0}=5 \times 10^{15}$ ${\rm ergcm}^{-2}{\rm s}^{-1}$.
          Solid wide line represents the case with conductivity, thin dotted
          line shows the pure radiative case.} 
\end{figure}
       
\subsection{Thermal instability criteria for the case with conductivity}

Now we would like to focus on the case with thermal conduction, because it 
seems 
to be more realistic. It is obvious that in medium with strong temperature 
gradient conductive heat transport plays a big role and influences 
the structure of 
region.   
In our model there are two very narrow (first $d\tau=1\times 10^{-4}$ and 
second $d\tau=1.6\times 10^{-6}$) zones with sharp temperature and density 
profiles in the transition layer between an accretion disc and a corona.
Thermal heat flux rises to  $3\times 10^8$ in these places.

To discuss the stability of such a solution we use local theory developed in
Section 4. As was mentioned in above this theory is valid for short period 
of time comparable with $1/Re(i\omega)$.
Otherwise we should treat 
our problem more globally which is more complicated and is not presented 
in this paper.
 
Another assumption of applicability of local theory is that  perturbation of 
considered parameter $x$ is smaller then characteristic scale length 
$x/\mid \nabla x \mid $ on which this parameter changes.      
Therefore we are looking for any perturbations bigger than 
$\lambda_{TC}$, but 
smaller than $T/\mid \nabla T \mid$, because they can
lead matter to condensate into clouds. 
Treating problem locally we determine 
the dependence of $\lambda_{TC}$ and the temperature scale length 
(named here $H_T$) on the optical depth. 

For a  wide range of $\tau$ the solution is stable, i. e.  $k_{TC}^2<0$ 
(equation (\ref{eq:ktc})).
Only for those two narrow zones with extremally strong temperature gradient,
wavenumber becomes positive and we can find size of perturbations that 
lead to thermal instability. 
Fig.3. presents variation of $\lambda_{TC}$ and $H_T$ versus $\tau$ for 
first (I) and second (II) zone, and for two different values of hard
radiation flux $F_{hard0}=5 \times 10^{14}$ a) and 
$F_{hard0}=5 \times 10^{15}$ b).
Because possible unstable regions are very thin, we rescale drawings by 
subtracting the optical depth of the beginning of zones $\tau_0$ 
from $\tau$.
Values of $\tau_0$ are written in the lower left corner for each case.
Note that the optical thickness $d\tau$ of zones does not depends on X-ray 
radiation flux.  
Only optical depth $\tau$ and distance of the zones  
from the disc surface $z$ 
is bigger for higher $F_{hard0}$. 

Characteristic temperature scale length is always one order of magnitude 
bigger than $\lambda_{TC}$ for the first zone in case aI) and more than
half of order of magnitude in case bI).
So the  perturbations with size $\lambda$ 
where $\lambda_{TC}< \lambda<H_T$ can lead to cloud condensation. 
The half order of magnitude difference between $\lambda_{TC}$ and $H_T$
is observed  for the second zone, but not from
the beginning where there is no place even for minimal perturbations 
(aII, bII).  
The higher hard radiation flux the smaller perturbations are required to 
produce thermal instability in the transition layer. 

\begin{figure}
 \epsfxsize = 120mm \epsfbox[50 400 530 700]{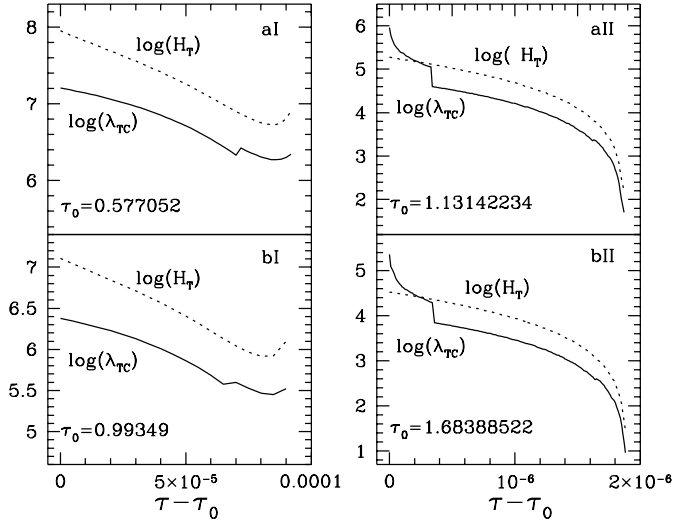}
 \caption{Characteristic length of perturbation $\lambda_{TC}$ 
          versus optical depth (solid line) compared with
          temperature scale length $H_{T}$  (dotted line) for first 
          (I) and second (II) unstable zone for two 
          different values of incident  hard radiation flux: a) 
          $F_{hard0}=5 \times 10^{14}$ ${\rm ergcm}^{-2}{\rm s}^{-1}$, b) 
          $F_{hard0}=5 \times 10^{15}$ ${\rm ergcm}^{-2}{\rm s}^{-1}$.
          $\tau_0$ is the optical depth of the beginning of zones.}
\end{figure}

%\begin{figure}
% \epsfxsize = 70 mm \epsfbox[30 160 490 700]{per.eps}
%
% \caption{Progress of $Re(i\omega)$ and $Im(i\omega)$ for particular 
%          optical depth $\tau=0.99354$ marked on Fig. 3bI. depending on 
%          size of perturbation $\lambda$. Long dashed line represents the 
%          $Re(i\omega)$ for different $\lambda$. Long dashed dotted line
%          shows $Im(i\omega)$ for different $\lambda$. Thin solid lines
%          mark critical size of perturbation $\lambda_{CT}$ and 
%          characteristic temperature scale length $H_T$ for $\tau=0.99354$.
%          Thin short dashed line marks $Re(i\omega)=0$.}  
%\end{figure}

We expect our clouds to be under influence of  gravity 
and particularly radiation pressure. The life time of clouds is very short
and probably they don't have enough time to grow up to the much bigger 
size than the minimal size of perturbations. So the most probable size 
of clouds is in order of $\lambda_{TC}$ and usually is taken into 
account in calculations of cloudy regions (Kurpiewski et al.1997,
Krolik 1998). As we can see in unstable zones there is not space for a
big clouds. But this is still enough for several clouds in 
size comparable  with  $\lambda_{TC}$ and because they are most important 
the results give good evidence for instability existence in the transition 
layer.  

\section{Discussion}

In this paper we described the influence of thermal conduction on the 
vertical structure of the transition layer between a hot corona and
a cool accretion disc. We found that heat transport via conductivity 
does not suppresses
thermal instability which occurs in the case when energy exchange 
through the disc/corona boundary is only due to radiation processes.

The temperature profile for the case with conductivity  is continuous 
but very sharp 
particularly in the place where thermal instability was found for pure 
radiative case. 
Local stability analysis of stationary state with thermal conduction
leads to the 
conclusion that in zones with extremely sharp temperature gradient,
the spontaneous perturbations with size $\lambda$, where $\lambda_{TC}<
\lambda<H_T$ can cause cloud condensation. The estimated size of possible 
clouds 
is $\sim \lambda$ and is much smaller than the size of thermally
unstable zone in pure radiative case which occurs $\sim 10^{10}$cm for 
both cases of $F_{hard0}$. 

Therefore we expect the structure from the corona surface to the disc
midplane as follows:
first the outermost zone, hot corona cooled by inverse Compton emission
(Begelman, Mckee \& Shields 1983, Begelman \& Mckee 1983, 
Ostricker, Mckee \& Klein 1991),
second, the transition layer with possibility of the development of the 
two-phase medium discussed in this paper,
and  third, main body of the disc (Shakura \& Sunnayev 1973).
We argue that upper parts of the 
transition consists of cool clouds at column density $N_H> 10^{22}$
embedded in hotter medium, but closer to the disc the situation changes 
continuously in the opposite one when 
hot blobs at $ N_H\sim 10^{19}$ are deep in cooler matter. 

Using local stability 
analysis we are not able to find the evolution of two-phase zones. We can
only find the possibility of clouds existence. 
We cannot to 
integrate full vertical disc structure, because ist is very difficult to 
describe all physical processes, particularly in the disc interior 
(three body recombination). 
Also on the distance from the Black Hole equal $10 r_{Sch}$ disc is 
radiation pressure dominated and another thermal  instability plays role.
Global analysis does not allow us to separate these two effects .
(In the disc interior convection may be important as well.) In addition we 
found that this is not  Sturm-Liouville problem, because we can find only 
finite number of eigenvalues (this is against main assumption 
of Sturm-Liouville problem).
In the future we should integrate equations numerically without any 
assumptions about character of perturbations and about differential 
operator. 

These results should be treated as preliminary and further research is 
necessary first to improve the physical content of the model and and 
second, 
to explore the range of parameters.
In the future work the transfer of hard radiation through the disc matter 
should be consider more carefully especially in the aim of better 
determination of absorption and reflection of X-rays depending on optical
depth. 

The position of possible unstable regions depends on the shape of 
atomic cooling and heating rates. 
We tested the accuracy of the adopted functions in our previous paper. 
Recent studies on photoionization codes 
show that abundance of iron and oxygen ions (with smaller contribution of 
magnesium, silicon and sulphur) plays crucial role in the 
nature of instability. Varying of this 
abundance strongly influences the heating and cooling behavior and the 
suppression of thermal instability might be achieved through small changes 
(Hess et al.1997).    

Our results are important in the aim of determine strictly the position of 
cloudy boundary layer.   
According to the present results, possible unstable regions in stationary 
state occur for $\Xi=0.28$ (first 
zone) and $\Xi=-0.19$ (second zone).Improved models should be used in the 
future to determine the dependence of these values on accretion disc 
parameters (including radius), X-ray radiation spectrum and chemical 
composition of the gas. The global stability analysis should be done 
by solving second order differention equation and finding global modes of
thermal instability.

We do not consider dissipative corona above transition layer. The heat 
transport via conductivity from such hot $\sim 10^9$K  region can change 
our picture. We expect to do it in the future.    

Unstable zones might be  very dynamical and it is  difficult to study 
the time evolution of such a region.
Clouds can evaporate or condense or to be accelerated by radiative pressure.
The first attempt to compute the spectrum from cloudy
transition layer was done by Kurpiewski et al. (1997). They found that 
broad emission lines observed in quasars  spectra can come from 
clouds which form continuously at the basis of the corona because of thermal
instability.

\section*{Acknowledgments}

Special thanks are due to Bo\. zena Czerny for her support and for many 
helpful conversations.
We thank Gary Ferland for providing us with his photoionization code
CLOUDY version 84.09. We are grateful to Wojciech Dziembowski,
Pawe{\l } Moskalik, Zbyszek Loska, Pawe{\l }  Magdziarz and
Leszek Zdunik for helpful discussions.
This work was supported in part 
by grant 2P03D00410 
of the Polish State Committee for 
Scientific Research.

\ \\
This paper has been processed by the authors using the Blackwell
Scientific Publications \LaTeX\  style file.

\end{document}